\documentclass[letterpaper,10pt,conference]{ieeeconf} 

\IEEEoverridecommandlockouts
\usepackage{times,amsmath,epsfig,amssymb,graphicx,amsfonts}
\usepackage{empheq}
\usepackage{graphicx}
\usepackage{psfrag}
\usepackage{fge}
\usepackage{amsmath}
\usepackage{setspace}
\usepackage{color}
\usepackage{amsfonts}   
\usepackage{amssymb}    
\usepackage{mathrsfs}
\usepackage{setspace}
\usepackage{dblfloatfix}
\usepackage{tikz}
\usepackage{tkz-tab}
\usepackage{algorithm}
\usepackage{algcompatible}
\usepackage{lipsum}
\usepackage{color}
\usepackage{mathrsfs}
\usepackage{epstopdf}
\usepackage{enumerate}
\usepackage{url}

\floatname{algorithm}{Algorithm}

\newtheorem{mydef}{Definition}
\newtheorem{myassump}{Assumption}
\newtheorem{mytheorem}{Theorem}
\newtheorem{mylemma}{Lemma}
\newtheorem{myremark}{Remark}
\newtheorem{mycorollary}{Corollary}

\newtheorem{myproblem}{Problem}
\newtheorem{mycontrolobj}{Objective}

\newcounter{ale}

\newenvironment{liste}{\begin{itemize}}{\end{itemize}}
\newcommand{\aliste}{\begin{liste} \setcounter{ale}{1}}
\newcommand{\zliste}{\end{liste}}

\title{{\LARGE {\bf Minimal Reachability Problems}}}

\author{V.~Tzoumas,  A.~Jadbabaie, G.~J.~Pappas{$^{\star}$}
\thanks{$^{\star}$All authors are with the Department of Electrical and Systems Engineering, University of Pennsylvania, Philadelphia, PA 19104-6228 USA (email: {\fontsize{8}{8}\selectfont\ttfamily\upshape \{vtzoumas, pappasg, jadbabai\}@seas.upenn.edu}).}
\thanks{This work was supported in part by TerraSwarm, one of six centers of STARnet, a Semiconductor Research Corporation program sponsored by MARCO and DARPA, in part by AFOSR Complex Networks Program and in part by ARO MURI W911NF-12-1-0509.}
}

\begin{document}
\maketitle

\begin{abstract}
In this paper, we address a collection of state space reachability problems, for linear time-invariant systems, using a minimal number of actuators.  In particular, we design a zero-one diagonal input matrix $B$, with a minimal number of non-zero entries, so that a specified state vector is reachable from a given initial state.  Moreover, we design a $B$ so that a system can be steered either into a given subset, or sufficiently close to a desired state.  This work extends the results of \cite{2013arXiv1304.3071O} and \cite{2014arXiv1401.4209P}, where a zero-one diagonal or column matrix $B$ is constructed so that the involved system is controllable. Specifically, we prove that the first two of our aforementioned problems are NP-hard; these results hold for a zero-one column matrix $B$ as well. Then, we provide efficient algorithms for their general solution, along with their worst case approximation guarantees.  Finally, we illustrate their performance over large random networks.
\end{abstract}


\section{Introduction}

Power grids, transportation systems, brain neural circuits and social networks are just a few of the complex dynamical systems that have drawn the attention of control scientists, \cite{amin2008electric,CalTransit, basset, mesbahi2010graph}, since their vast size, and interconnectivity, necessitate novel control techniques with regard to:
\begin{enumerate}[i.]
\item tasks that are collective \cite{murray2007recent}, e.g., reaching consensus in a system of autonomous interacting vehicles \cite{1673588};
\item new cost constraints, e.g., with respect to the number of used actuators and the level of the input and communication power \cite{6781658}.
\end{enumerate}

In this paper, we consider a set of minimal state reachability problems, for linear time-invariant systems, where the term `minimal' captures our objective to use the least number of actuators towards the involved control tasks.  Specifically, we design a zero-one diagonal input matrix $B$, with a minimal number of non-zero entries, so that one of the following (collective) tasks are met: i) the resultant system can be steered into a subset, or ii) to a state, or iii) sufficiently close to a state.
Therefore, our work relaxes the objective of \cite{2013arXiv1304.3071O} and \cite{2014arXiv1401.4209P}, where a zero-one diagonal or column matrix $B$ is constructed, with a minimal number of non-zero entries, so that the designed system is controllable.  

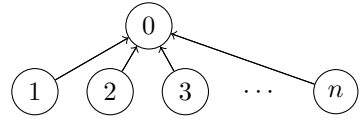
\begin{figure}[t]
\centering
\begin{tikzpicture}
\tikzstyle{every node}=[draw,shape=circle];
\node (v0) at (0:0) {$1$};
\node (v1) at (0:1) {$2$};
\node (v2) at (0:2) {$3$};
\node[white] (v5) at (0:3) {\color{black} $\cdots$};
\node (v3) at (0:4) {$n$};
\node (v4) at (30:1.75) {$0$};

\foreach \from/\to in {v0/v4, v1/v4, v2/v4, v3/v4}
\draw [->] (\from) -- (\to);
\draw
(v0) -- (v4)
(v1) -- (v4)
(v2) -- (v4)
(v3) -- (v4);

\end{tikzpicture}
\caption{A $n+1$-node star network: each node represents a state of a linear time-invariant system of the form $dx/dt=Ax+Bu$ (where $x=(x_0, x_1, \ldots, x_n)$ is the state vector; $A$ is the system's matrix; $B$ is the input matrix; and $u$ is the input vector).  The state of node `0' depends on the states of all the nodes in the network.}
\label{fig:star}
\end{figure}

This is an important distinction whenever we are interested only in the feasibility of a state transfer, as in power grids \cite{amin2008electric}; transportation systems \cite{CalTransit}; complex neural circuits \cite{basset}; infection processes over large-scale social networks \cite{2013arXiv1309.6270P} (e.g., from the infectious state to the state where all the network nodes are healthy): Consider for example the system in Fig.~\ref{fig:star} and assume the transfer from the initial state zero to $(1, 0, 0, \ldots, 0)$, where the first entry corresponds to the final state of node `0', the second to that of `1', and so forth; if we impose controllability in the design of $B$, we get a $B$ with $n$ non-zero elements: $B=\text{diag}(0,1,1,\ldots,1)$; that is, states $x_1$ through $x_n$ must be actuated so that this system is controllable.  On the other hand, if we impose only state reachability, we get a $B$ with only one non-zero element, independently of $n$; e.g., a solution is $B=\text{diag}(1,0,0,\ldots,0)$, where only state $x_0$ is actuated.  Thereby, whenever we are interested in the feasibility of a state transfer and in a $B$ with a small number of non-zero elements, the objective of state reachability should not be substituted with that of controllability: under controllability the number of used actuators could grow linearly with $n$, while under state reachability it could be one for all $n$.  Similar comments carry through with respect to the rest of our objectives.

At the same time, the task to design a sparsest zero-one diagonal matrix $B$ is combinatorial, and, as a result, it may be computationally hard in the worst case.  Indeed, we prove that the first two of our aforementioned problems are NP-hard --- our proofs hold for a zero-one column matrix $B$ as well.  Therefore, we then provide efficient algorithms for their general solution, along with their worst case approximation guarantees; to this end, we use an approximation algorithm that we provide for our third problem, where a sparse zero-one diagonal matrix $B$ is designed so that a system can be steered $\epsilon$-close to a desired state.

These hardness results proceed by reduction to the minimum hitting set problem (MHS), which is NP-hard \cite{arora2009computational}.  In particular, we prove that the problem of state reachability, using a minimal number of actuators, is NP-hard, by reducing it to the controllability problem introduced in \cite{2013arXiv1304.3071O}, which is at least as hard as the MHS.  Moreover, we prove that the problem of steering a system into a subset is NP-hard by directly reducing it to the MHS.

Then, we first provide an efficient approximation algorithm so that a system can be steered $\epsilon$-close to a desired state.
This algorithm returns a $B$ with a number of non-zero elements up to a multiplicative factor of $O(\ln(\epsilon^{-1}))$ from any optimal solution.  
Therefore, 
it allows the designer to select the level of approximation $\epsilon$, with respect to the trade-off between the reachability error $\epsilon$ and the number of used actuators (recall that the number of non-zero elements of $B$ coincides with the number of used actuators).  Afterwards, we use this algorithm to provide efficient approximation algorithms for the rest of our reachability problems as well.

In addition to \cite{2013arXiv1304.3071O} and \cite{2014arXiv1401.4209P}, other relevant studies to this paper are \cite{2014arXiv1404.7665S,bullo2014,acc2015} and \cite{2014arXiv1409.3289T}, where their authors consider the design of a sparse input matrix $B$ so that an input energy objective is minimized.  Moreover, \cite{dhingra2014admm} and \cite{6882815} address the sparse design of the closed loop linear system, with respect to its feedback gain, as well as, a set of sensor placement problems.  Other recent works that study sensor placement problems are the \cite{joshi2009sensor} and \cite{6701125}. 

Furthermore, \cite{blondel2000survey} considers the decidability of a set of problems related to ours; for example, it asks whether the problem of deciding if there exists a control that can drive a given system from an initial state to a desired one is decidable or not.  
The main difference between this set of problems and ours is that they consider the feasibility of state transfer given a fixed system, whereas we design a system so that the feasibility of a state transfer is guaranteed. 

The remainder of this paper is organized as follows.  The formulation and model for our reachability problems are set forth in Section \ref{sec:pform}, where the corresponding integer optimization programs are stated. In Section \ref{sec:np}, we prove the intractability of these problems and, then, in Section \ref{sec:alg}, we provide efficient algorithms for their general solution, along with their worst case approximation guarantees.  Finally, in Section \ref{sec:ex}, we illustrate our analytical findings, using an instance of the network in Fig.~\ref{fig:star}, and afterwards, we test the efficiency of the proposed algorithms over large random networks that are commonly used to model real-world networked systems.  Section \ref{sec:con} concludes the paper.

\section{Problem Formulation}\label{sec:pform}

\paragraph*{Notation}
We denote the set of natural numbers $\{1,2,\ldots\}$ as $\mathbb{N}$, the set of real numbers as  $\mathbb{R}$,  and we let $[n]\equiv \{1, 2, \ldots, n\}$ for all $n \in \mathbb{N}$.  Also, given a set $\mathcal{X}$, we denote as $|\mathcal{X}|$ its cardinality.  Matrices are represented by capital letters and vectors by lower-case letters. For a matrix ${A}$, ${A}^{T}$ is its transpose and $A_{ij}$ is its element located at the $i-$th row and $j-$th column. 
Moreover, we denote as ${I}$ the identity matrix; its dimension is inferred from the context.  Additionally, for ${\delta} \in \mathbb{R}^n$, we let $\text{diag}({\delta})$ denote an $n \times n$ diagonal matrix such that $\text{diag}({\delta})_{ii}=\delta_i$ for all $i \in [n]$.  The rest of our notation is introduced when needed.

\subsection{Model}

Consider a linear system of $n$ states, $x_1, x_2,\ldots,x_n$, whose evolution is described by
\begin{align}
\dot{{x}}(t) = {A}{{x}}(t) + {B}{{u}}(t), t > t_0,
\label{eq:dynamics}
\end{align} where  $t_0 \in \mathbb{R}$ is fixed, ${x}\equiv \{x_1,x_2,\ldots,x_n\}$, $\dot{{x}}(t)\equiv d{x}/dt$, and ${u} \in \mathbb{R}^n$ is the input vector.  The matrices ${A}$ and ${B}$ are of appropriate dimension. Without loss of generality, ${u} \in \mathbb{R}^n$; in general, whenever the $i$-th column of $B$ is zero, $u_i$ is ignored.  Moreover, we denote \eqref{eq:dynamics} as the duple $(A,B)$ and refer to the states $x_1, x_2,\ldots, x_n$ as nodes $1, 2,\ldots, n$, respectively; finally, we denote their collection as $\mathcal{V}\equiv[n]$.

In what follows, ${A}$ is fixed and the following structure is assumed on $B$:

\begin{myassump}\label{assump:Diag_B}
$B$ is a diagonal zero-one matrix:
${B}=\emph{\text{diag}}({\delta})$, where ${\delta}\in\{0,1\}^{n}$.
\end{myassump}

Therefore, if $\delta_i=1$, state $x_i$ is actuated, and if $\delta_i=0$, is not and $u_i$ is ignored.  That is, the number of non-zero elements of $B$ coincides with the number of actuators (inputs) that are implemented for the control of system \eqref{eq:dynamics}.

In this paper, we design $B$ so that $(A,B)$ satisfies a control objective among the following presented in the next section.

\subsection{Minimal Reachability Problems}

We introduce two control objectives, the \textit{state} and \textit{subset reachability}, which we use to define the design problems of this paper.  In particular, consider $t_0$, $t_1 \geq t_0$, and $x(t_0)$ fixed:

\begin{mycontrolobj}[State Reachability]\label{obj:1}
The state $\chi \in \mathbb{R}^n$ is reachable by $(A,B)$ at time $t=t_1$ if and only if there exists input defined over $(t_0,t_1)$ such that $x(t_1)=\chi$.
\end{mycontrolobj}

A parallel notion to the state reachability is the \textit{state feasibility}: 

\begin{mydef}[State Feasibility]\label{def:state_feas}
The transfer from $x(t_0)$ to $x(t_1)=\chi \in \mathbb{R}^n$ by $(A,B)$, denoted as $x(t_0) \rightarrow x(t_1)=\chi$, is feasible if and only if $\chi$ is reachable by $(A,B)$ at time $t=t_1$.
\end{mydef}

We now present our second objective:

\begin{mycontrolobj}[Subset Reachability]\label{obj:2}
The subset $\mathcal{N} \subseteq \mathbb{R}^n$ is reachable by $(A,B)$ at time $t=t_1$ if and only if there exist $\chi\in \mathcal{N}$ and input defined over $(t_0,t_1)$ such that $x(t_1)=\chi$ is reachable.
\end{mycontrolobj}


The corresponding definition of \textit{subset feasibility} parallels that of state feasibility and it is omitted. 

Evidently, Objective \ref{obj:2} generalizes Objective \ref{obj:1}: According to it, $(A,B)$ targets  from $x(t_0)$ a subset, instead of a single state.  Nevertheless, subset reachability of $\mathcal{N}$ does not imply that all states $\chi \in \mathcal{N}$ are reachable.  Similarly, although $\chi \in \mathcal{N}$ may not be reachable by $(A,B)$, $\mathcal{N}$ can be; thus, Objective \ref{obj:1} is not a special case of Objective \ref{obj:2}.  Overall, Objectives \ref{obj:1} and \ref{obj:2} define the two separate design problems that follow.

\begin{myproblem}[Minimal State Reachability] \label{pr:1}
Given $x(t_0)$ and $x(t_1)$, design a $B$ with the smallest number of non-zero elements so that the state transfer $x(t_0) \rightarrow x(t_1)$ is feasible.
\end{myproblem}
Note that Problem \ref{pr:1} is always feasible, since for any $A$, $(A,I)$ is controllable.

Therefore, the objective of Problem \ref{pr:1} relaxes that of \cite{2013arXiv1304.3071O,2014arXiv1401.4209P} where $B$ is designed  with the smallest number of non-zero elements so that the resultant $(A,B)$ is controllable.  

\begin{myproblem}[Minimal Subset Reachability]\label{pr:2} Given $x(t_0)$, $\mathcal{N}$ and $t_1$, design a $B$ with the smallest number of non-zero elements so that the subset $\mathcal{N}$ is reachable from $x(t_0)$ at time $t_1$.
\end{myproblem}
We refer to Problem \ref{pr:2} as \textit{minimal subset reachability} as well. As with Problem \ref{pr:1}, Problem \ref{pr:2} is always feasible, since for any $A$, $(A,I)$ is controllable.

Evidently, the `minimal' term in the definition of Problems \ref{pr:1} and \ref{pr:2} captures our objective to design a sparsest\footnote{A matrix is \textit{sparse} if it has a small number of non-zero elements compared to each dimension.} $B$. 

%

Finally, all of our results carry through if we consider the output  $y(t)=Wx(t)$ of \eqref{eq:dynamics}, where $W$ is fixed and of appropriate dimension, instead of $x(t)$.  In particular, denote as $\mathcal{R}(W)$ the column space of $W$ and consider the following objectives:

\begin{mycontrolobj}[Output Reachability]\label{obj:3}
The output state $y \in \mathcal{R}(W)$ is reachable by $(A,B)$ at time $t=t_1$ if and only if there exists input defined over $(t_0,t_1)$ such that $y(t_1)=y$.
\end{mycontrolobj}

Naturally, Objectives \ref{obj:1} and \ref{obj:3} coincide for $W=I$. Thereby, a generalized version of Problem \ref{pr:1}, where a sparsest $B$ is designed so that an output transfer is feasible, is due.  Similar comments apply with respect to the objective below.

\begin{mycontrolobj}[Output Subset Reachability]\label{obj:4}
The $\mathcal{N} \subseteq$ $\mathcal{R}(W)$ is reachable by $(A,B)$ at time $t=t_1$ if and only if there exist $y \in \mathcal{N}$ and input defined over $(t_0,t_1)$ such that $y(t_1)=y$ is reachable.
\end{mycontrolobj}

In what follows, we continue with the original Problems \ref{pr:1} and \ref{pr:2}.

\section{Main Results}\label{sec:main}

In the first part of this section, \ref{sec:np}, we prove that Problems \ref{pr:1} and \ref{pr:2} are NP-hard.  The proofs proceed by reduction to the minimum hitting set problem (MHS), which is NP-hard \cite{arora2009computational}, and is defined as follows:

\begin{mydef}[Minimum Hitting Set Problem] \label{def:MHS} Given a finite set $\mathcal{M}$ and a collection $\mathcal{L}$ of non-empty subsets of $\mathcal{M}$, find a smallest cardinality $\mathcal{M'}\subseteq \mathcal{M}$ that has a non-empty intersection with each set in $\mathcal{L}$.
\end{mydef}

In particular, we prove that Problem \ref{pr:1} is NP-hard providing an instance that reduces to the controllability problem introduced in \cite{2013arXiv1304.3071O}, which is at least as hard as the MHS; as a result, we conclude that Problem \ref{pr:1} is as well.  Moreover, we prove that Problem \ref{pr:2} is NP-hard by directly reducing it to the MHS.

In the second part of this section, \ref{sec:alg}, since Problems \ref{pr:1} and \ref{pr:2} are NP-hard, we provide efficient approximation algorithms for their general solution.  Towards this direction, we first generalize Definition \ref{def:state_feas} as follows:

\begin{mydef}[$\epsilon$-close feasibility]\label{def:e-feas}
The transfer $x(t_0) \rightarrow x(t_1)=\chi \in \mathbb{R}^n$ by $(A,B)$ is $\epsilon$-feasible if and only if there exists $\chi' \in \mathbb{R}^n$ reachable by $(A,B)$ at time $t=t_1$ such that $\|\chi-\chi'\|^2\leq \epsilon$, where $\|\cdot\|$ denotes the euclidean norm.
\end{mydef}

For $\epsilon=0$, Definitions \ref{def:state_feas} and \ref{def:e-feas} coincide.

We use Definition \ref{def:e-feas} to relax the objective Problem \ref{pr:1}, by replacing the feasibility of $x(t_0)\rightarrow x(t_1)$ with that of $\epsilon$-close feasibility --- from a real-world application perspective, and for small $\epsilon$, this is a weak modification: the convergence of a system exactly to a desired $x(t_1)$ is usually infeasible, e.g., due to external disturbances.  We then provide for this problem a polynomial time approximation algorithm, Algorithm \ref{alg:min_e_feas}, that returns a $B$ with sparsity\footnote{The \textit{sparsity} of a matrix is the number of its non-zero elements.} up to a multiplicative factor of $O(\ln(\epsilon^{-1}))$ from any optimal solution of the original Problem \ref{pr:1}.  


Next, to address Problem \ref{pr:1} with respect to Objective \ref{obj:1}, we prove that for all $\epsilon \leq \epsilon(A)$, where $\epsilon(A)$ is positive and sufficiently small, Definitions \ref{def:state_feas} and \ref{def:e-feas} still coincide; hence, we implement a bisection-type execution of Algorithm \ref{alg:min_e_feas}, Algorithm \ref{alg:min_feas}, that quickly converges to an $\epsilon \leq \epsilon(A)$ and, as a result, returns a $B$ that makes the exact transfer $x(t_0) \rightarrow x(t_1)$ feasible.  

Finally, we provide an approximation algorithm for Problem \ref{pr:2} when $\mathcal{N}\subseteq \mathbb{R}^n$ is finite, by observing that in this case $\mathcal{N}$ can be approximated as a finite union of euclidean balls in $\mathbb{R}^n$.  Specifically, let $\chi_1, \chi_2, \ldots, \chi_{k(\mathcal{N})}$ be their centres and $\epsilon_1, \epsilon_2, \ldots, \epsilon_{k(\mathcal{N})}$ their corresponding radii.  Moreover, without loss of generality, assume $x(t_0)=0$.  Then, by executing Algorithm \ref{alg:min_e_feas} for $(x(t_1)=\chi_i, \epsilon =\epsilon_i)_{i \in [k(\mathcal{N})]}$ and selecting the sparsest solution $B$ among all $i \in [k(\mathcal{N})]$, we return an approximate solution to Problem \ref{pr:2} with Algorithm's \ref{alg:min_e_feas} worst case guarantees.


\subsection{Intractability of the Minimal Reachability Problems}\label{sec:np}

We prove that Problems \ref{pr:1} and \ref{pr:2} are NP-hard.  
The proofs proceed with respect to the decision version of Problems \ref{pr:1} and \ref{pr:2} and that of MHS.  The latter is defined as follows:

\begin{mydef}[$k$-hitting set]\label{def:k_HS}
Given a finite set $\mathcal{M}$ and a collection $\mathcal{L}$ of non-empty subsets of $\mathcal{M}$, find an $\mathcal{M'}\subseteq \mathcal{M}$ of cardinality at most $k$ that has a non-empty intersection with each set in $\mathcal{L}$.
\end{mydef}
Without loss of generality, we assume that every element of $\mathcal{M}$ appears in at least one set in $\mathcal{L}$ and all set in $\mathcal{L}$ are non-empty.

The decision versions of Problems \ref{pr:1} and \ref{pr:2} are defined in Sections \ref{sec:np1} and \ref{sec:np2}, where we present their NP-hardness, respectively.

\subsubsection{Intractability of Problem \ref{pr:1}}\label{sec:np1}

We prove that the decision version of Problem \ref{pr:1} reduces to the $k$-hitting set and, as a result, that Problem \ref{pr:1} is NP-hard.

This version of Problem \ref{pr:1} is defined by replacing the feasibility objective with that of $k$-feasibility:

\begin{mydef}[$k$-feasibility]\label{def:state_k_feas}
The transfer $x(t_0)\rightarrow x(t_1)$ is $k$-feasible if and only if there exists  $k$-sparse\footnote{A matrix is \textit{$k$-sparse} if it has $k$ non-zero elements.} $B$ such that $x(t_0)\rightarrow x(t_1)$ is feasible by $(A,B)$.
\end{mydef}

To present our instance of the decision Problem \ref{pr:1} that reduces to the $k$-hitting set problem, let $|\mathcal{L}|=p$ and $\mathcal{M}=\{1, 2,$ $\ldots, m\}$, with respect to Definition \ref{def:k_HS}, and define $\Phi \in \mathbb{R}^{p \times m}$ such that $\Phi_{ij}=1$ if the $i$-th set contains the element $j$ and zero otherwise.

\begin{mylemma}\label{th:feas_contr_of_same_size}
For $i \in \mathbb{N}$, denote as $e_{i\times l}$ the $i\times l$ matrix of all-ones and set $n = m+p+1$, $A = V_1^{-1} \text{diag}(1,2,\ldots,$ $m+p+1) V_1$, where\footnote{$V_1$ is invertible since it strictly diagonally dominant.}
\[ 
V_1= \left[ 
\begin{array}{ccc}
2I_{m\times m} & 0_{m \times p} & e_{m\times 1} \\
\Phi & (m+1)I_{p\times p} & 0_{p \times 1} \\
0_{1 \times m} & 0_{1 \times p} & 1
\end{array}\right],
\]
and $x(t_0)=0$, as well as, $\chi=V_1^{-1}e_{n\times 1}$.
For any $t_1>t_0$, $0\rightarrow x(t_1)=\chi$ is $k+1$-feasible if and only if $\mathcal{L}$ has a $k$-hitting set.
\end{mylemma}

Therefore, with Lemma \ref{th:feas_contr_of_same_size} we provide an instance of Problem \ref{pr:1} that is $k+1$-feasible if and only if any instance of $\mathcal{L}$, (that is, also the hardest ones with respect to the hitting set problem), has a $k$-hitting set. Hence (cf.~\cite{arora2009computational}): 

\begin{mytheorem}\label{th:pr1_np}
Problem \ref{pr:1} is NP-hard.
\end{mytheorem}

Thereby, the generalized version of Problem \ref{pr:1}, with respect to Objective \ref{obj:3}, is NP-hard as well (for the above instance where we additionally set $W=I$).

We illustrate the proof Lemma \ref{th:feas_contr_of_same_size}: The instance of $A$ and the initial and final condition are constructed so that the $0 \rightarrow \chi$ is $k+1$-feasible if and only if there exists $k+1$-sparse $B$ such that $(A,B)$ is controllable; on the other hand, the latter holds if and only if $\mathcal{L}$ has a $k$-hitting set \cite{2013arXiv1304.3071O}.  Thereby, the theorem follows.  Additionally, due to the controllability properties of linear time-invariant systems \cite{Chen:1998:LST:521603}, it holds for any $t_1 > t_0$.

However, the proof of Lemma \ref{th:feas_contr_of_same_size} suggests that the sparse reachability of a system is hard merely because its sparse controllability is.  To show the contrary, we generalize Lemma \ref{th:feas_contr_of_same_size} by constructing an $A$ and a $x(t_0)\rightarrow x(t_1)$ so that $x(t_0)\rightarrow x(t_1)$ is $k+1$-feasible if and only if $\mathcal{L}$ has a $k$-hitting set, while the resultant system is not controllable.

\begin{mylemma}\label{th:feas_contr_of_diff_size}
For $i \in \mathbb{N}$, denote as $e_{i\times l}$ the $i\times l$ matrix of all-ones and set $n=m+p+2$, $A= V_2^{-1}\text{diag}(1,2,\ldots,$ $m+p+2)V_2$, where
\[ 
V_2= \left[ 
\begin{array}{cccc}
2I_{m\times m} & 0_{m \times p} & e_{m\times 1} & 0_{m \times 1}\\
\Phi & (m+1)I_{p\times p} & 0_{p \times 1} & 0_{p \times 1}\\
0_{1 \times m} & 0_{1 \times p} & 1 & 0 \\
0_{1\times m} & 0_{1\times p} & 0 & 1
\end{array}\right],
\]
and $x(t_0)=0$, as well as, $\chi=V_2^{-1}\left[ 
\begin{array}{c}
e_{1\times (n-1)}, 0
\end{array}\right]^T$.
For any $t_1>t_0$, the $x(t_0)\rightarrow x(t_1)=\chi$ is $k+1$-feasible if and only if $\mathcal{L}$ has a $k$-hitting set.
\end{mylemma}

With this instance, we prove that $0 \rightarrow \chi$ is $k+1$-feasible if and only if a sub-system of $(A,B)$ is $k+1$-controllable, a fact that is equivalent to $\mathcal{L}$ having a $k$-hitting set \cite{2013arXiv1304.3071O}.  On the other hand, $(A,B)$ remains uncontrollable.  Therefore, the NP-hardness of Problem \ref{pr:1} emanates from this class of instances as well, where state reachability is achieved without implying controllability to the resultant system.

Lemma \ref{th:feas_contr_of_same_size} extends to the case where $B$ is a column zero-one vector as well.  Furthermore, in Theorem \ref{th:pr1_np} the assumption $x(t_0)=0$ is without loss of generality, since we consider the linear dynamics \eqref{eq:dynamics} \cite{Chen:1998:LST:521603}.
Finally, Lemmas \ref{th:feas_contr_of_same_size} and \ref{th:feas_contr_of_diff_size} extend to the case where $B$ is a column zero-one vector as well.  Furthermore, in both theorems, the assumption $x(t_0)=0$ is without loss of generality, since we consider the linear dynamics \eqref{eq:dynamics} \cite{Chen:1998:LST:521603}.

In the following paragraphs, we prove the NP-hardness of Problem \ref{pr:2}.

\subsubsection{Intractability of Problem \ref{pr:2}}\label{sec:np2}

We prove that the decision version of Problem \ref{pr:2} reduces to the $k$-hitting set and, as a result, that Problem \ref{pr:2} is NP-hard.

This version of Problem \ref{pr:2} is defined by replacing the reachability objective with that of $k$-reachability:

\begin{mydef}[$k$-reachability]\label{def:k_reach}
The subset $\mathcal{N} \subseteq \mathbb{R}^n$ is $k$-reachable if and only if there exists $k$-sparse $B$ such  that $\mathcal{N}$ is reachable by $(A,B)$.
\end{mydef}

To present our instance of the decision Problem \ref{pr:2} that reduces to the $k$-hitting set problem, let $|\mathcal{L}|=p$ and $\mathcal{M}=\{1, 2,$ $\ldots, m\}$, with respect to Definition \ref{def:k_HS}, and define $\Phi\in \mathbb{R}^{p \times m}$ such that $\Phi_{ij}=1$ if the $i$-th set contains the element $j$ and zero otherwise.

\begin{mylemma}\label{th:NP_reach}
Set $\mathcal{N}=\{(x_1,x_2,\ldots,x_n): x_1=x_2=\ldots=x_m=0, x_{m+1}, x_{m+2},\ldots, x_{m+p} >0\}$ and
\[ 
A= \left[ 
\begin{array}{cc}
0_{m\times m} & 0_{m \times p} \\
\Phi & 0_{p \times p}  
\end{array}\right].
\]
$\mathcal{N}$ is $k$-reachable if and only if $\mathcal{L}$ has a $k$-hitting set.
\end{mylemma}

Therefore, with Lemma \ref{th:NP_reach} we provide an instance of Problem \ref{pr:2} that is $k$-feasible if and only if any instance of $\mathcal{L}$, (that is, also the hardest ones with respect to the hitting set problem), has a $k$-hitting set. Hence (cf.~\cite{arora2009computational}): 

\begin{mytheorem}
Problem \ref{pr:2} is NP-hard.
\end{mytheorem}

Thereby, the generalized version of Problem \ref{pr:2}, with respect to Objective \ref{obj:4}, is NP-hard as well (for the above instance where we additionally set $W=I$).

Since Problems \ref{pr:1} and \ref{pr:2} are NP-hard, we need in the worst case to provide approximate algorithms for their solution; this is the subject of the next section.

\subsection{Approximation Algorithms for the Minimal Reachability Problems}\label{sec:alg}

We provide efficient approximation algorithms for the general solution of Problems \ref{pr:1} and \ref{pr:2}.  
Recall that these problems aim for a sparse $B$ so that a  transfer is feasible or a subset of the state space is reachable, respectively. At the same time, the sparsity of $B$ equals the number of actuators that we should implement in system \eqref{eq:dynamics} so to satisfy these goals.  Therefore, the objective of these algorithms is the sparse control of system \eqref{eq:dynamics}.

To implement an approximation algorithm for Problem \ref{pr:1}, we use Definition \ref{def:e-feas} to relax Objective \ref{obj:1}, by replacing the feasibility of $x(t_0)\rightarrow x(t_1)$ with that of $\epsilon$-close feasibility.  We then provide Algorithm \ref{alg:min_e_feas}, that returns a $B$ with sparsity up to a multiplicative factor of $O(\ln(\epsilon^{-1}))$ from any optimal solution of the original Problem \ref{pr:1}.

Next, to address Problem \ref{pr:1} with respect to Objective \ref{obj:1}, we prove that for all $\epsilon \leq \epsilon(A)$, where $\epsilon(A)$ is positive and sufficiently small, Definitions \ref{def:state_feas} and \ref{def:e-feas} still coincide; hence, we implement a bisection-type execution of Algorithm \ref{alg:min_e_feas}, Algorithm \ref{alg:min_feas}, that quickly converges to an $\epsilon \leq \epsilon(A)$ and, as a result, returns a $B$ that makes the exact transfer $x(t_0) \rightarrow x(t_1)$ feasible. 

Finally, using Algorithm \ref{alg:min_e_feas}, we provide an approximation algorithm for Problem \ref{pr:2} as well.

\subsubsection{Approximation Algorithm for Problem \ref{pr:1}}\label{sec:alg1}

We develop the notation and tools that lead to an efficient approximation algorithm for Problem \ref{pr:1}. 

For $\mathcal{N}\subseteq\mathbb{R}^n$ and $v \in \mathbb{R}^{n\times 1}$, we denote as $v[\mathcal{N}]$ the projection of $v$ onto $\mathcal{N}$ and as $\|v\|$ its euclidean norm.  Moreover, we denote as $\mathcal{C}(A)$ the set of columns of $\left[I|A|\ldots|A^{n-1}\right]$, as $e_i$ the $i$-th unit vector and as $C_i$ the set of columns $\{e_i, Ae_i,\ldots, A^{n-1}e_i\}$.  For $B$ per Assumption~\ref{assump:Diag_B}, we set
\[
\mathcal{S}(B)\equiv\text{span}\left[B|AB|\ldots|A^{n-1}B\right].
\]

Since the dynamics \eqref{eq:dynamics} are linear, $x(t_0)\rightarrow x(t_1)$ is feasible if and only if $0 \rightarrow  x(t_1)-\exp[A(t_1-t_0)]x(t_0)\equiv v(t_1)$ is.  Moreover, since these dynamics are also continuous and time-invariant, whenever $0 \rightarrow v(t_1)$ is feasible for some $t_1 > t_0$, it is also for any $t_1' > t_0$ \cite{Chen:1998:LST:521603}.  Hence, we study directly $0 \rightarrow v$, suppressing $t_1$.

In particular, $0 \rightarrow v$ is feasible if and only if $v \in \mathcal{S}(B)$ \cite{Chen:1998:LST:521603}.  Therefore, $0 \rightarrow v$ is feasible if and only if $v=v[\mathcal{S}(B)]$:  if $v=v[\mathcal{S}(B)]$, $v \in \mathcal{S}(B)$, while, if $v\neq v[\mathcal{S}(B)]$, $v-v[\mathcal{S}(B)] \in \mathcal{S}(B)^\perp$, that is, $v \notin \mathcal{S}(B)$\footnote{$\mathcal{S}(B)^\perp$ is the orthogonal complement of $\mathcal{S}(B)$.}.  Similarly, $0 \rightarrow v$ is feasible if and only if $\|v\|=\|v[\mathcal{S}(B)]\|$: if $v=v[\mathcal{S}(B)]$, $\|v\|=\|v[\mathcal{S}(B)]\|$, while, if $v\neq v[\mathcal{S}(B)]$, $\|v[\mathcal{S}(B)]\| < \|v\|$.

Definition \ref{def:e-feas} is restated as follows:

\begin{mydef}[$\epsilon$-close feasibility]\label{def:e-feas_2}
The $0 \rightarrow v$ is \emph{$\epsilon$-close feasible} by $(A,B)$ if and only if  $\|v\|^2-\|v[\mathcal{S}(B)]\|^2 \leq \epsilon$.
\end{mydef}

\begin{myremark}
Since $v-v[\mathcal{S}(B)]$ is orthogonal to $v[\mathcal{S}(B)]$, $\|v[\mathcal{S}(B)]\|^2+\|v-v[\mathcal{S}(B)]\|^2=\|v\|^2$ and, as a result, $\epsilon$-close feasibility implies $\|v-v[\mathcal{S}(B)]\|^2\leq \epsilon$.
\end{myremark}

We provide the following greedy approximation algorithm for Problem \ref{pr:1} with respect to the relaxed feasibility objective of Definition \ref{def:e-feas_2}. Its quality of approximation is quantified in Theorem \ref{th:alg_1}.

\begin{algorithm}
\caption{Approximation Algorithm for the relaxed Problem \ref{pr:1} with respect to Definition \ref{def:e-feas_2}.}\label{alg:min_e_feas}
\begin{algorithmic}
\REQUIRE Matrix $\mathcal{C}(A)$, vector $v \equiv x(t_1)-\exp[A(t_1-t_0)]x(t_0)$, approximation level $\epsilon$.
\ENSURE $B$ such that $x(t_0)\rightarrow x(t_1)$ is $\epsilon$-close feasible.
\STATE $B=0_{n\times n}$.
\WHILE {$\|v\|^2-\|v\left[\mathcal{S}(B)\right]\|^2 > \epsilon$} 
\STATE{ 	
\mbox{} Find an $i \in [n]$ such that: i) $B_{ii}=0$ and ii) $i$ is a maximizer for $\|v\left[\mathcal{S}(B)+\text{span}\{C_i\}\right]\|^2- \|v\left[\mathcal{S}(B)\right]\|^2$.

Set $B_{ii}=1$.
	}
\ENDWHILE
\end{algorithmic} 
\end{algorithm}

\begin{mytheorem}\label{th:alg_1}
Given the transfer $x(t_0)\rightarrow x(t_1)$, denote as $B^\star$ an optimal solution to Problem \ref{pr:1} and as $B$ the corresponding output of Algorithm \ref{alg:min_e_feas}.  Then, $x(t_0)\rightarrow x(t_1)$ is $\epsilon$-close feasible by $(A,B)$ and 
\[
\sum\limits_{i=1}^{n}B_{ii}\leq \lceil\ln(\| x(t_1)-\exp[A(t_1-t_0)]x(t_0)\|^2/\epsilon)\rceil\sum\limits_{i=1}^{n}B^\star_{ii}.
\]
\end{mytheorem}
That is, the polynomial time approximation Algorithm \ref{alg:min_e_feas} returns a $B$ with sparsity up to a multiplicative factor of $O(\ln(\epsilon^{-1}))$ from any optimal solution of the original Problem \ref{pr:1}, and makes the $x(t_0)\rightarrow x(t_1)$, or $0 \rightarrow v$, $\epsilon$-close feasible.


Next, to address Problem \ref{pr:1} with respect to Objective \ref{obj:1}, we show that there exists $\epsilon(A)$, positive, such that for any $\epsilon \leq \epsilon(A)$, Definitions \ref{def:state_feas} and \ref{def:e-feas} coincide.  Thereby, running Algorithm \ref{alg:min_e_feas} with $\epsilon \leq \epsilon(A)$, results to a $B$ that makes the exact transfer $x(t_0) \rightarrow x(t_1)$ feasible.

In particular, for $i\in [n]$, let $C_i\equiv \{e_i, Ae_i,\ldots, A^{n-1}e_i\}$; that is, $C_i$ is the sub-matrix of $\mathcal{C}(A)$ that is also present in $\left[B|AB|\ldots|A^{n-1}B\right]$ if and only if $B_{ii}=1$.  Moreover, for $S\subseteq [n]$, consider $B_{ii}=1$ if and only if $i \in S$.  Moreover, assume that $0 \rightarrow v$ is infeasible by $B$, i.e., $v[\text{span}\{\bigcup_{j\in S} C_j\}]\neq v$.  Then, denote as $\Xi(S)$ the event where $0 \rightarrow v$ can become feasible by making one more element of $B$ one, that is,  $\Xi(S)\equiv \{v[\text{span}\{\bigcup_{j\in S} C_j\}]\neq v \text{ and } \exists i \in [n] \setminus S, v[\text{span}\{\left(\bigcup_{j\in S} C_j\right)\cup C_i\}]=v\}$.  It is, 
\[
\epsilon(A)=\min_{S\subseteq [n]: \Xi(S) \text{ is true.}}\left(\|v\|^2-\|v[S]\|^2\right).
\]
Therefore, $\epsilon(A)$ is positive.  

 
In general, $\epsilon(A)$ is unknown in advance.  Hence, we need to search for a sufficiently small value of $\epsilon$ so that $\epsilon \leq \epsilon(A)$.  Since $\epsilon$ is lower and upper bounded by $0$ and $\|v\|^2$, respectively, we achieve this by performing a binary search.  In particular, we implement Algorithm \ref{alg:min_feas}, where we denote as  $[\text{Algorithm} \ref{alg:min_e_feas}](\mathcal{C}(A), 0\rightarrow v, \epsilon)$ the matrix that Algorithm~\ref{alg:min_e_feas} returns for given $A$, $v$ and $\epsilon$.


\begin{algorithm}
\caption{Approximation Algorithm for Problem \ref{pr:1}.}\label{alg:min_feas}
\begin{algorithmic}
\REQUIRE Matrix $\mathcal{C}(A)$, vector $v\equiv x(t_1)-\exp[A(t_1-t_0)]x(t_0)$,  bisection's accuracy level $a$.
\ENSURE $B$ such that $x(t_0)\rightarrow x(t_1)$ is feasible.
\STATE $B=0_{n\times n}$, $l\leftarrow 0$, $u\leftarrow \|v\|^2$, $\epsilon\leftarrow(l+u)/2$
\WHILE {$u-l>a$}\\	
    $B \leftarrow [\text{Algorithm} \ref{alg:min_e_feas}](\mathcal{C}(A), 0\rightarrow v, \epsilon)$
	\IF {$\|v\|^2-\|v\left[\mathcal{S}(B)\right]\|^2 > \epsilon$} \STATE{$u\leftarrow \epsilon$} \ELSE \STATE{$l\leftarrow \epsilon$} 
	\ENDIF\\
	$\epsilon\leftarrow (l+u)/2$   
\ENDWHILE
\IF {$\|v\|^2-\|v\left[\mathcal{S}(B)\right]\|^2 > \epsilon$} \STATE{$u\leftarrow \epsilon$}, $\epsilon\leftarrow (l+u)/2$
\ENDIF\\
$B \leftarrow [\text{Algorithm} \ref{alg:min_e_feas}](\mathcal{C}(A), 0\rightarrow v, \epsilon)$
\end{algorithmic} 
\end{algorithm}

In the worst case, when we first enter the \texttt{while} loop, the \texttt{if} condition is not satisfied and, as a result, $\epsilon$ is set to a lower value.  This process continues until the \texttt{if} condition is satisfied for the first time, from which point and on, the algorithm converges, up to the accuracy level $a$, to $\epsilon(A)$; specifically, $|\epsilon-\epsilon(A)| \leq a/2$, due to the mechanics of the bisection.  Then, Algorithm~\ref{alg:min_feas} exits the \texttt{while} loop and the last \texttt{if} statement ensures that $\epsilon$ is set below $\epsilon(A)$ so that $0\rightarrow v$ is feasible. 

The efficiency of Algorithm \ref{alg:min_feas} for Problem \ref{pr:1} is summarized below.

\begin{mycorollary}\label{th:alg_2}
Given the transfer $x(t_0)\rightarrow x(t_1)$, denote as $B^\star$ an optimal solution to Problem \ref{pr:1} and as $B$ the corresponding output of Algorithm \ref{alg:min_feas}.  Then, $x(t_0)\rightarrow x(t_1)$ is feasible by $(A,B)$ and 
\[
\sum\limits_{i=1}^{n}B_{ii}\leq \lceil\ln(\| x(t_1)-\exp[A(t_1-t_0)]x(t_0)\|^2/\epsilon)\rceil\sum\limits_{i=1}^{n}B^\star_{ii}.
\]
where $\epsilon$ is the approximation level where Algorithm \ref{alg:min_feas} had converged when terminated.
\end{mycorollary} 

The results of this section apply to the generalized version of Problem \ref{pr:1} with respect to Objective \ref{obj:3} by replacing  $\mathcal{C}(A)$, $C_i$ and $\mathcal{S}(B)$ with $W\mathcal{C}(A)$, $WC_i$ and $\text{span}\left[WB|WAB|\ldots|WA^{n-1}B\right]$, respectively (where $W$ is the output matrix of \eqref{eq:dynamics}).  Similarly with regard to the approximation algorithm described below.

\subsubsection{Approximation Algorithm for Problem \ref{pr:2}}\label{sec:alg2}

We sketch the approximation algorithm for Problem \ref{pr:2} (for the case where $\mathcal{N}\subseteq \mathbb{R}^n$ is finite), since, then, its implementation is straightforward: Without loss of generality, assume $x(t_0)=0$, as the dynamics \eqref{eq:dynamics} are linear, and consider the problem of reaching a finite $\mathcal{N}\subseteq \mathbb{R}^n$.  Observe that $\mathcal{N}$ can be approximated as a finite union of euclidean balls in $\mathbb{R}^n$.  Specifically, let $\chi_1, \chi_2, \ldots, \chi_{k(\mathcal{N})}$ be their centres and $\epsilon_1, \epsilon_2, \ldots, \epsilon_{k(\mathcal{N})}$ their corresponding radii.  Then, by executing Algorithm \ref{alg:min_e_feas} for $(\mathcal{C}(A), 0\rightarrow \chi_i, \epsilon =\epsilon_i)_{i \in [k(\mathcal{N})]}$ and, afterwards, selecting the sparsest solution $B$ among all $i \in [k(\mathcal{N})]$, we return an approximate solution to Problem \ref{pr:2}.  As in Algorithm \ref{alg:min_e_feas}, two levels of approximation underlie here:  First, we approximate $\mathcal{N}$ with a sufficient number of balls, and, then, we approximate the sparsity of the optimal solution to Problem \ref{pr:2}; the quality of the latter approximation is quantified in Theorem \ref{th:alg_1}.

We illustrate our analytical findings, and test their performance, in the next section.

\section{Examples and Discussions}\label{sec:ex}

We test the performance of Algorithm \ref{alg:min_feas} over various systems, starting in Subsection \ref{subsec:star} with the networked system of Fig.~\ref{fig:star} and following up in Subsection~\ref{subsec:randomGraphs} with Erd\H{o}s-R\'{e}nyi random networks. 
Extending the simulations of this section to the algorithm for Problem \ref{pr:2} is straightforward and, as a result, due to space limitations we omit this discussion.

\subsection{Star Network}\label{subsec:star}

We illustrate the mechanics and efficiency of Algorithm \ref{alg:min_feas} using the star network of Fig.~\ref{fig:star}, where $n=4$ and
\begin{align}
{A} = \left[\begin{array}{ccccc}
-1 & 1 & 1 & 1 & 1\\
0 & -1 & 0 & 0 & 0\\
0 & 0 & -1 & 0 & 0\\
0 & 0 & 0 & -1 & 0\\
0 & 0 & 0 & 0 & -1
\end{array}\right].\nonumber
\end{align}

In particular, we run Algorithm \ref{alg:min_feas} for the $\tau_1 \equiv 0\rightarrow (1,0,0,0,0)$, $\tau_2 \equiv0\rightarrow (0,1,1,0,0)$ and $\tau_3 \equiv0\rightarrow (1,1,1,0,0)$ and for $a=.001$. The algorithm returned a $B$ equal to $\text{diag}(1,0,0,0,0)$, $\text{diag}(0,1,1,0,0)$ and $\text{diag}(0,1,1,0,0)$, respectively; indeed, $\tau_1$ is feasible by the minimum number of actuators if and only if either $x_0(t)$ is actuated or one among $x_1(t),x_2(t),x_3(t),x_4(t)$ is; $\tau_2$ is feasible by the minimum number of actuators if and only if $x_1(t)$ and $x_2(t)$ are actuated and, finally, $\tau_3$ is feasible by the minimum number of actuators if and only if $x_1(t)$ and $x_2(t)$ are actuated.  Overall, Algorithm \ref{alg:min_feas} operated optimally.

Evidently, this star network is controllable by the minimum number of actuators if and only if all $x_1(t),x_2(t),x_3(t),x_4(t)$ are actuated.  Therefore, whenever we are interested merely in the feasibility of a state transfer, it is cost-effective, with respect to the number of actuators that should be implemented, to design a $B$ that does not result to a controllable system as well. 

\subsection{Erd\H{o}s-R\'{e}nyi Random Networks}\label{subsec:randomGraphs}

Erd\H{o}s-R\'{e}nyi random graphs are commonly used to model real-world networked systems~\cite{newman2006structure}.  According to this model, each edge is included in the generated graph with some probability $p$ independently of every other edge.  We implemented this model for varying network sizes $n$ where the directed edge probabilities were set to $p = 2\log(n)/n$.  In particular, we first generated the binary adjacencies matrices for each network size so that each edge is present with probability $p$ and then we replaced every non-zero entry with an independent standard normal variable to generate a randomly weighted graph.  The network size varied from $1$ to $100$, with step $1$.  
 
For each network size, we run Algorithm \ref{alg:min_feas} for a $0 \rightarrow \chi$, where $\chi$ was randomly generated using MATLAB's ``randn'' command; for all cases, the algorithm returned a $1$-sparse $B$.  This is in accordance with the simulation results of \cite{2013arXiv1304.3071O}, where similarly randomly generated networks were made controllable by actuating one or two states.

Extending the simulations of this section to the algorithm for Problem \ref{pr:2} is straightforward and, as a result, due to space limitations we omit this discussion.

\section{Concluding Remarks}\label{sec:con}

We addressed a collection of state (and output) space reachability problems for a linear system, under the additional objective of sparse control, i.e., the control using a minimal number of actuators.  
In particular, we proved that these problems are NP-hard
and provided efficient approximation algorithms for their general solution, along with worst case approximation guarantees.  Finally, we illustrated the efficiency of these algorithms with a set of simulations.  Optimal behaviour was observed.

Moreover, any optimal control problem, e.g., the LQR. where an objective is optimized with respect to i) the input vector $u$ and ii) the sparsity of $B$, subject to the system dynamics, as well as, an initial and final condition of the form $x(t_0)\in \mathbb{R}^n $ and $x(t_1)\in \mathbb{R}^n$ or $x(t_1)\in \mathcal{N} \subseteq \mathbb{R}^n$, respectively, is NP-hard as well. This conclusion suggests a future direction: Which is an efficient approximation algorithm for such optimal control problems?  A relevant result is \cite{acc2015}, where the authors provide an efficient approximation algorithm for minimizing the input energy for a desired state transfer, subject to a $k$-sparse $B$ and a controllable $(A,B)$.

Finally, due to Lemmas \ref{th:feas_contr_of_same_size} and \ref{th:NP_reach}, and since for the hitting set problem it is NP-hard to find a set whose cardinality is within a factor of $O(\log(n))$ from the optimal set \cite{moshkovitz2012projection}, it is an open problem to find for Problem \ref{pr:1} an approximation algorithm that achieves an $O(\log(n))$ approximation factor, or to prove that this is the case for Algorithm \ref{alg:min_feas}.



\appendices

\section{Proofs of the Main Results}\label{proofs}

\subsection{Lemma \ref{th:feas_contr_of_same_size}}

\begin{proof}
Denote as $r_i$ the $i$-th row of $V_1$. It is proved in \cite{2013arXiv1304.3071O} that $\mathcal{L}$ has a $k$-hitting set if and only if $A$ is $k+1$-controllable (that is, $(A,B)$ is controllable for $B$ being $k+1$-sparse).  Therefore, we prove that $0 \rightarrow \chi$ is $k$-feasible at time $t_1$ by $(A,B)$ if and only if $A$ is $k$-controllable.

If $0 \rightarrow \chi$ is $k$-feasible at time $t_1$, then 
\[
\chi=\int\limits_{t_0}^{t_1}e^{A(t_1-\tau)}Bu(\tau)d\tau,
\]
for some input $u$ defined over $(t_0,t_1)$.  Let $\epsilon\equiv \epsilon(t_1)$ such that $e^{n(t_1-t_0)}\leq 1+\epsilon$ and observe that all the entries of $A$ are non-negative.  Then,
\[
e_{n\times 1}\leq(1+\epsilon)V_1B\int\limits_{t_0}^{t_1}u(\tau)d\tau.
\]
Set $v\equiv\int\limits_{t_0}^{t_1}u(\tau)d\tau$.  Therefore, $e_{n\times 1}\leq (1+\epsilon)V_1Bv$:  Assume that there exists $i$ such that $r_iB=0$. Then, $r_i B v=0 < 1$; contradiction.  As a result, for all $i\in [n]$, $r_iB\neq 0$, which implies, from the PBH theorem, that $A$ is $k$-controllable.

Conversely, if $A$ is $k$-controllable, then $0 \rightarrow \chi$ is $k$-feasible at any time $t>t_0$ by $A$, that is, also for $t=t_1$.
\end{proof}

\subsection{Lemma \ref{th:feas_contr_of_diff_size}}

Due to space limitations, this proof is omitted; it can be found in the full version of this paper, located at the authors websites.
\subsection{Lemma \ref{th:NP_reach}}

\begin{proof}
Let $\mathcal{P}\equiv \{(x_1,x_2,\ldots,x_n): x_1=x_2=\ldots=x_m=0, x_{m+1}, x_{m+2},\ldots, x_{m+p} >0\}$,

Assume that $\mathcal{S}$ is a hitting set of cardinality at most $k$ for $\mathcal{L}$.  For all $i\in\mathcal{S}$, set $B_{ii}=1$.  Then, there exists $\chi \in \mathcal{P}$, $\chi \in \text{span}\{[B | AB]\}$, i.e., $\mathcal{P}$ is $k$-reachable, since by writing $B$ as
\[ 
B= \left[ 
\begin{array}{cc}
B(1)_{m\times m} & 0_{m \times p} \\
0_{p \times m} & B(2)_{p \times p}  
\end{array}\right],
\]
then
\[
[B | AB]=\left[\begin{array}{cccc}
B(1)_{m\times m} & 0_{m \times p} & 0 & 0\\
0_{p \times m} & B(2)_{p \times p} & \Phi B(1)_{m\times m} & 0  
\end{array}\right].
\]

Conversely, assume that $\mathcal{P}$ is $k$-reachable.  That is, there exists $\chi \in \mathcal{P}$, $\chi \in \text{span}\{[B | AB]\}$ and consider $[B | AB]$: Choose an $i$ such that $B(2)_{ii}=1$ and the smallest $j \in [m]$ such that $\Phi_{ij}=1$: Set $B(2)_{ii}=0$ and $B(1)_{jj}=1$.  It remains true that there exists $\chi' \in \mathcal{P}$ (possibly different than $\chi$), $\chi' \in \text{span}\{[B | AB]\}$, i.e., that $\mathcal{P}$ is $k$-reachable. Proceeding likewise for all $i$ such that $B(2)_{ii}=1$, we construct a $k$-sparse matrix $B(1)$, (while $B(2)$ becomes zero).  Then, the set $\{j: B(1)_{jj}=1\}$ is a $k$-hitting set for $\mathcal{L}$.
\end{proof}

\subsection{Theorem \ref{th:alg_1}}

\begin{proof}
We denote as $\mathcal{I}$ a set of columns of $\mathcal{C}(A)$ such that $v\left[\text{span}\{\cup_{c \in \mathcal{I}}c\}\right]=v$ and the cardinality of $\mathcal{I}(\#)\equiv \left\{i:\exists c \in \mathcal{I}, c\in \{e_i, Ae_i,\ldots, A^{n-1}e_i\}\right\}$ is minimum.  Also, we denote as $B(\mathcal{I})$ the zero-one diagonal matrix such that $B_{ii}(\mathcal{I})=1$ if and only if $i\in \mathcal{I}(\#)$.  That is, $B(\mathcal{I})$ is a sparsest matrix such that $0 \rightarrow v$ is feasible.

For any $S\subseteq \mathcal{C}(A)$,
\[
v\left[\text{span}\{S\cup_{i\in \mathcal{I}(\#)}C_i\}\right]=v.
\]
As $i$ successively runs over all the elements of $\mathcal{I}(\#)$, $\|v\|^2-\|v[\text{span}\{S\cup \cdot\}]\|^2$ decreases from $\|v\|^2-\|v[\text{span}\{S\}]\|^2$ to $0$.  Thereby, there is some $i'$ for which the dimension decreases by at least $(\|v\|^2-\|v[\text{span}\{S\}]\|^2)/|\mathcal{I}(\#)|$; otherwise, the total decrease is strictly less that $\|v\|^2-\|v[\text{span}\{S\}]\|^2$, contradiction.  Thus, denoting as $\mathcal{I}(\#)\setminus i'$ the previous indices of $i'$ in the succession,
\begin{align*}
\|v\|^2-\|v\left[\text{span}\{(S\cup_{i\in \mathcal{I}(\#)\setminus i'}C_i) \cup C_{i'}\}\right]\|^2\leq\\
\|v\|^2-\|v\left[\text{span}\{S\cup_{i\in \mathcal{I}(\#)\setminus i'}C_i\}\right]\|^2\\
-\frac{\|v\|^2-\|v[\text{span}\{S\}]\|^2}{|\mathcal{I}(\#)|}.
\end{align*}
Furthermore, from Lemma 8.1 in \cite{sviridenko2014optimal}
\begin{align*}
\|v\left[\text{span}\{S\cup (C_{i'}\setminus S)\}\right]\|^2-\|v\left[\text{span}\{S\}\right]\|^2 \geq\\ \|v\left[\text{span}\{(S\cup_{i\in \mathcal{I}(\#)\setminus i'}C_i)\cup  (C_{i'}\setminus (S\cup_{i\in \mathcal{I}(\#)\setminus i'}C_i))\}\right]\|^2\\- \|v\left[\text{span}\{S\cup_{i\in \mathcal{I}(\#)\setminus i'}C_i\}\right]\|^2,
\end{align*}
and since $\text{span}\{S\cup (C_{i'}\setminus S)\}=\text{span}\{S\cup C_{i'}\}$ and $\text{span}\{(S\cup_{i\in \mathcal{I}(\#)\setminus i'}C_i)\cup  (C_{i'}\setminus ((S\cup_{i\in \mathcal{I}(\#)\setminus i'}C_i)))\}=\text{span}\{(S\cup_{i\in \mathcal{I}(\#)\setminus i'}C_i)\cup  C_{i'})\}$, 
\begin{align}\label{ineq:basic_alg_1}
\|v\|^2-\|v\left[\text{span}\{S\cup C_{i'}\}\right]\|^2  \leq \\
\left(1-\frac{1}{|\mathcal{I}(\#)|}\right) \left(\|v\|^2-\|v[\text{span}\{S\}]\|^2\right).\label{ineq:basic_alg_2}
\end{align}

At Algorithm~\ref{alg:min_e_feas}, consider that the \texttt{while} loop has been executed for $k$ times, and let $B_k$ denote the corresponding constructed matrix.  By the inequality in \eqref{ineq:basic_alg_1}-\eqref{ineq:basic_alg_2}, there is an $i$ such that the next time that the \texttt{while} loop will be executed
\begin{align*}
\|v\|^2-\|v\left[\mathcal{S}(B_{k+1})\right]\|^2  \leq \\
\left(1-\frac{1}{|\mathcal{I}(\#)|}\right) \left(\|v\|^2-\|v[\mathcal{S}(B_k)]\|^2\right).
\end{align*}
Thus,
\begin{align*}
\|v\|^2-\|v\left[\mathcal{S}(B_{k+1})\right]\|^2 \leq \ldots \leq \\
\left(1-\frac{1}{|\mathcal{I}(\#)|}\right)^{k}\|v\|^2 \leq 
e^{-k/|\mathcal{I}(\#)|}\|v\|^2.
\end{align*}

Thereby, after $\bar{k}\equiv |\mathcal{I}(\#)|\lceil\ln(\|v\|^2/\epsilon)\rceil$ steps (with $|\mathcal{I}(\#)|$ being equal to the number of the non-zero elements of $B(\mathcal{I})$),
\begin{align*}
\|v\|^2-\|v\left[\mathcal{S}(B_{\bar{k}})\right]\|^2 \leq \epsilon,
\end{align*}
and, as a result, $0 \rightarrow v$ is $\epsilon$-close feasible.
\end{proof}

\bibliographystyle{IEEEtran}
\bibliography{newRef}

\end{document}